\documentclass[twocolumn,secnumarabic,amssymb, nobibnotes, aps, prd,nofootinbib]{revtex4-2}

\usepackage{graphicx}
\usepackage{caption}
\usepackage{subcaption}
\usepackage{xcolor}
\usepackage{hyperref}

\makeatletter
\def\blfootnote{\xdef\@thefnmark{}\@footnotetext}
\makeatother

\newcommand{\ode}[2]{\frac{\mathrm{d} #1}{\mathrm{d} #2}} 
\newcommand{\pde}[2]{\frac{\partial #1}{\partial #2}}

\begin{document}
\received{4 June 2020}
\revised{18 March 2021}
\accepted{19 March 2021}
\published{13 August 2021}

%\title{Detonation model using Burgers equation and a pulsed reaction}%
\title{On the two separate decay time scales of a detonation wave modelled by the Burgers equation and their relation to its chaotic dynamics}

\author{S. S.-M. \surname{Lau-Chapdelaine} \href{https://orcid.org/0000-0003-1541-6183}{\includegraphics[keepaspectratio,width=0.7em]{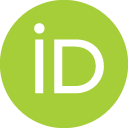}}}%
\email[Contact: \href{mailto:shem.lau-chapdelaine@rmc.ca}{shem.lau-chapdelaine@rmc.ca}; currently at: Department of Chemistry and Chemical Engineering, Royal Military College,
Kingston, Ontario, Canada, K7K 7B4]{}
\affiliation{Department of Mechanical Engineering, University of Ottawa, 
	Ottawa, Ontario, Canada, K1N 6N5}
\author{M. I. Radulescu \href{https://orcid.org/0000-0002-2752-9313}{\includegraphics[keepaspectratio,width=0.7em]{orcid.png}}}%
\affiliation{Department of Mechanical Engineering, University of Ottawa, 
	Ottawa, Ontario, Canada, K1N 6N5}
%\date{\today}%

\begin{abstract}
This study uses a simplified detonation model to investigate the behaviour of detonations with galloping-like pulsations. The reactive Burgers equation is used for the hydrodynamic equation, coupled to a pulsed source whereby all the shocked reactants are simultaneously consumed at fixed time intervals. The model mimics the short periodic amplifications of the shock front followed by relatively lengthy decays seen in galloping detonations.
Numerical simulations reveal a saw tooth evolution of the front velocity with a period-averaged detonation speed equal to the Chapman-Jouguet velocity. The detonation velocity exhibits two distinct groups of decay time scales, punctuated by reaction pulses.
At each pulse, a rarefaction wave is created at the reaction front's last position. A characteristic investigation reveals that characteristics originating from the head of this rarefaction take 1.57 periods to reach and attenuate the detonation front, while characteristics at the tail take an additional period. The leading characteristics are amplified twice, by passing through the reaction fronts of subsequent pulses, before arriving at the shock front, whilst the trailing characteristics are amplified three times. This leads to the two distinct groups of time scales seen in the detonation front speed.
\end{abstract}

\blfootnote{Originally published in Phys. Rev. E 104, 025103. \\ DOI: \href{https://doi.org/10.1103/PhysRevE.104.025103}{10.1103/PhysRevE.104.025103}.\\ This authors' version was compiled \today.\\\hrule}

\maketitle
%\tableofcontents

\section{Introduction}
Recently, Radulescu and Tang \cite{radulescu_nonlinear_2011} and Kasimov \textit{et al.} \cite{kasimov_model_2013} have found that the reactive Burgers equation describing one-dimensional detonation waves in reactive media admits chaotic pulsating dynamics, and follows the classical period-doubling route to chaos seen in many other non-linear systems. The reactive Burgers equation recovers the more general observations of chaotic dynamics from detonations modeled with the Euler \cite{ng_nonlinear_2005,henrick_simulations_2006} or Navier-Stokes equations \cite{romick2012effect}. Their discovery of chaotic dynamics relied on numerical integration of the partial differential equations but did not explain why the system follows that universal route.

The periodic solution of pulsating detonations can be characterized in two parts, one associated with the very rapid re-amplification of the detonation by energy release behind the front, and the second by a long inert-like decay of the lead shock. This has been observed experimentally in so-called ``galloping'' detonations in thin tubes \cite{he1995dynamical,jackson_detonation_2016}, and numerically in low-velocity detonations \cite{sow2017stabilization}. 

\begin{figure*}[]
	\centering
	\begin{subfigure}[t]{0.24\textwidth}
		\centering
		\includegraphics[width=1.06\textwidth]{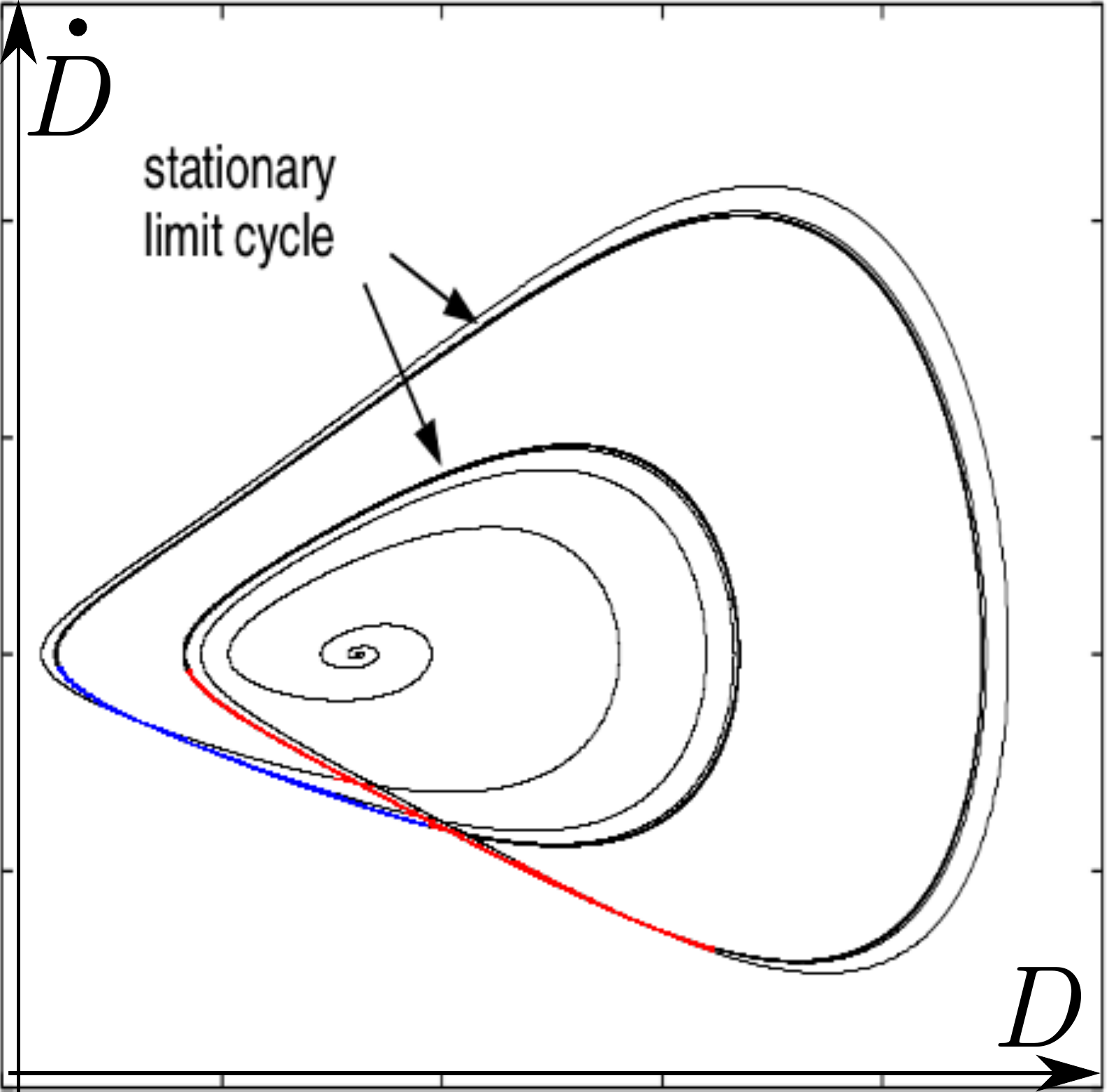}
		%\caption{\label{fig:Henrick_et_al_2006_fig7}Euler equations \cite{henrick_simulations_2006}}
		\caption{}
		\label{fig:Henrick_et_al_2006_fig7}
	\end{subfigure}\hfill
	\begin{subfigure}[t]{0.25\textwidth}
		\centering
		\includegraphics[width=0.905\textwidth]{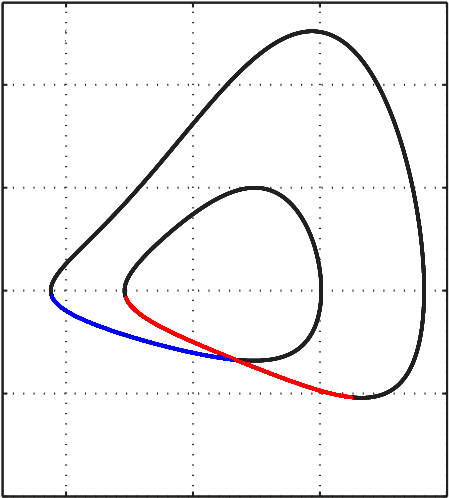}
		%\caption{\label{fig:Kasimov_2013_fig2}Asymptotic \cite{kasimov_model_2013}}
		\caption{}
		\label{fig:Kasimov_2013_fig2}
	\end{subfigure}%
	\begin{subfigure}[t]{0.5\textwidth}
		\centering
		\includegraphics[width=0.493\textwidth]{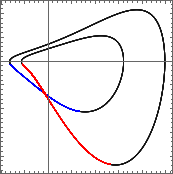}%
		\label{fig:period_two:Bellerive_et_al} %\hfill%
		\includegraphics[width=0.495\textwidth]{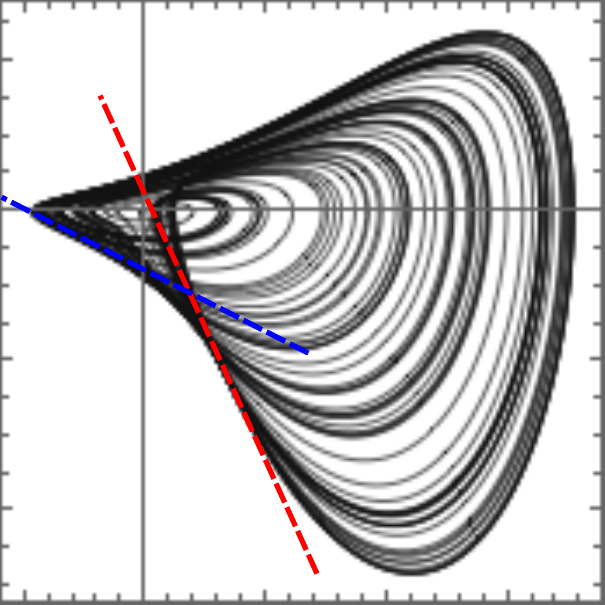}%
		%\caption{\label{fig:Bellerive_2015_chaos}Evolution equation from Fickett's model \cite{bellerive_chaos_2015}}
		\caption{}
		\label{fig:Bellerive_2015_chaos}
	\end{subfigure}%
	\caption{{Phase diagrams of period-two detonations (three left subfigures) and a chaotic detonation (right) in three models: (a) the Euler equations \cite{henrick_simulations_2006}, (b) an asymptotic model of the Euler equations \cite{kasimov_model_2013}, and (c) an evolution equation derived from Fickett's detonation analogue \cite{bellerive_chaos_2015}; two groups of decay rates are highlighted in red and blue}}
	\label{fig:period_two}
\end{figure*}

Nevertheless, the chaotic dynamics shown in the limit cycle of figure \ref{fig:period_two} are characterized by three time scales. There are two distinct groups of decay of the lead shock, highlighted in red and blue, punctuated by rapid re-amplification periods. 
%Limit cycles of period-two oscillations are shown in figure \ref{fig:period_two} for detonations in the Euler equations and two simplified models. These phase diagrams have two distinct groups of decay rates highlighted in red and blue, separated by re-amplifications. 
Each decay rate appears as a straight line with constant slope $\dot{D}/D$ in figure \ref{fig:period_two} and represents a characteristic time scale ($D$ is the detonation velocity and $\dot{D}$ is its time derivative).
These two distinct groups of decay time scales seem to be intrinsic to the period-doubling bifurcations, are inherited by subsequent period-doubling bifurcations, and persist into chaos. The two decay groups are observed in the Euler equations (figure \ref{fig:Henrick_et_al_2006_fig7}) and Burgers-based models (figures \ref{fig:Kasimov_2013_fig2} and \ref{fig:Bellerive_2015_chaos}).

The present study seeks to explore why there are two distinct periods of decay in chaotic detonations. In order to focus on the decay time scales only, a pulsating model is assumed in which the re-amplification stage is infinitely fast and occurs periodically. This is based on a model introduced by Radulescu and Shepherd \cite{radulescu2015dynamics} for the reactive Euler equations. In their model, inert hydrodynamics of shock decay were periodically interrupted by the instantaneous release of all chemical energy stored in the unreacted gas accumulated behind the lead shock. The sudden pressure gain caused by reactions was followed by the inert decay of the shock front and the shock wave traveling into the products of the previous pulse

A similar model was studied by Mi and Higgins using the Euler equations \cite{miEffectSpatialDiscretization2017} and the Burgers equation \cite{mi_influence_2015}. In their model, a shock wave traveled through an inert medium interleaved with thin, regularly spaced sources of energy. After a prescribed time delay, a shocked source would release all of its energy into the flow, causing a local blast wave behind the shock front, initiating the re-amplification phase. This phase terminated when the forward-traveling portion of the blast reached and amplified the front. The decay phase then began once again.

Motivated by the period-doubling bifurcation route to chaos, this paper introduces a similarly simple model to study the two distinct groups of decay rates. The Euler equations are simplified to the Burgers equation, and the fast dynamics of re-amplification are replaced by simple pulsations in order to focus on decay time scales. This combination of simplifications was presented by Lau-Chapdelaine and Radulescu \cite{lau-chadeplaine_detonation_2019}. %, a pulsating model is assumed in which the re-amplification stage is infinitely fast and occurs periodically. 
In contrast to Mi and Higgins' work \cite{mi_influence_2015,miEffectSpatialDiscretization2017}, the entire medium is reactive and reactions are periodically forced in time, not at discrete points in space. At every pulse time, all shocked yet unburnt gas in the detonation is instantaneously reacted. This effectively makes the re-amplification phase infinitely fast, leaving only the dynamics of decay. The decay dynamics are studied numerically and analytically to find the source of the two distinct groups of decay timescales seen in chaotic detonations.

The model is described in further detail in section \ref{sec:model} and the numerical method is explained in section \ref{numerical_method}. Simulation results are presented in section \ref{sec:results} and an analytical model is developed and tested against the simulations in section \ref{sec:analysis}. The conclusion can be found in section \ref{sec:conclusion}.

\section{\label{sec:model}Model}

{\color{black}The Burgers equation with a source term}
\begin{equation}
\pde{u}{t} + \frac{1}{2}\pde{}{x}\left(u^2 + \lambda q \right) = 0
\label{eq:reactive_Burgers_equation}
\end{equation}
{\color{black}is used for the hydrodynamic equation. The variable $u$ is the local information speed, $t$ is time, and $x$ is location. %, but is also analogous to density \cite{fickett_detonation_1979}.
The hydrodynamics are coupled to a source term $q \lambda$ to account for reactions. The constant $q=1$ is the heat released by combustion and the reaction progress variable $\lambda=0$ when unburnt, or $\lambda=1$ when burnt.}

{\color{black}The reactive Burgers equation arises naturally in transonic flows with weak energy release. This occurs, for example, %in modelling of condensed-phased explosives 
in detonations where most of the energy is released quickly, followed by a slower release of the remainder near a sonic plane \cite{bdzilTheoryMachReflection2017}. The equation can also be asymptotically derived \cite{rosales_weakly_1983,clavin_dynamics_2002,faria_theory_2015} from the Navier-Stokes equations in the Newtonian limit with weak heat release. }

%The hydrodynamic equation{\color{black}, the reactive Burgers equation, arises naturally in transonic flows with weak energy release, like the  \cite{bdzilTheoryMachReflection2017}. It} can also be asymptotically derived {\color{black}\cite{rosales_weakly_1983,clavin_dynamics_2002,faria_theory_2015}} from the Navier-Stokes equations in the Newtonian limit with weak heat release. The Burgers equation with a source term is
%\begin{equation}
%\pde{u}{t} + \frac{1}{2}\pde{}{x}\left(u^2 + \lambda q \right) = 0
%\label{eq:reactive_Burgers_equation}
%\end{equation}
%where $t$ is time, and $x$ is location. The variable $u$ is the local information speed. %, but is also analogous to density \cite{fickett_detonation_1979}.
%The hydrodynamics are coupled to a source term $q \lambda$ to account for reactions. The constant $q=1$ is the heat released by combustion and the reaction progress variable $\lambda=0$ when unburnt, or $\lambda=1$ when burnt.

Burgers-based detonation analogues were first introduced by Fickett \cite{fickett_detonation_1979} and Majda \cite{majda_qualitative_1981} as qualitative models for reactive gas dynamics problems. They have since been used to understand a range of dynamic detonations phenomena including direct initiation \cite{lau-chapdelaine_planar_2017}, the eigenvalue structure and limits in the presence of losses \cite{fickett_introduction_1985,faria2015qualitative}, their stability \cite{clavin_dynamics_2002,radulescu_nonlinear_2011,kasimov_model_2013,kabanov2018linear}, {\color{black}glancing detonation reflections \cite{bdzilTheoryMachReflection2017}}, detonations in heterogeneous systems \cite{mi_influence_2015,lau-chadeplaine_multiplicity_2019}, and rotating detonation engines \cite{kochModelockedRotatingDetonation2020}. The reactive Burgers equation (\ref{eq:reactive_Burgers_equation}) was used to model the hydrodynamics in these studies with reaction models that differed between applications.

A simple reaction model that captures the slow decay of the detonation velocity below the steady Chapman-Jouguet (CJ) velocity followed by a rapid re-amplification \cite{he1995dynamical,jackson_detonation_2016,sow2017stabilization} is sought by taking the limit of an infinitely fast reamplification compared to the decay phase.
This is accomplished by forcing reactions at fixed time intervals $t_{\mathrm{p}}=1$, independent of the hydrodynamics. At each pulse, all shocked reactants are consumed completely, instantly, and simultaneously. 
No reactions occur between pulses and the ``reaction front'', the interface between burnt and unburnt gasses, remains stationary in the model's frame of reference (referred to as the ``laboratory'' frame of reference herein). In other words, whenever the simulation time $t$ reaches a multiple $n$ of the pulse time $t_{\mathrm{p}}$, $\lambda$ is set to one to the left of the shock, and $\lambda$ remains zero to the right of the shock; \textit{i.e.}, $\lambda =	1$ $\forall$ $x \le x_{\mathrm{s}}$ when $t = n t_{\mathrm{p}}$, for $n \in \mathbb{N}$; $x_{\mathrm{s}}$ is the shock position. The resulting step-like reaction profile is equivalent to having a reaction zone that is very thin compared to the detonation structure.

The pulsed reaction greatly simplifies the complicated dynamics of reamplification of unsteady detonations. Using the Burgers equation also simplifies the characteristics, leaving only forward-traveling characteristics which move at the speed $u$ to be considered.

The problem is scaled by parameters $q$ and $t_{\mathrm{p}}$, so values of 1 are used for both. The Rankine-Hugoniot jump conditions for this system (shown later in equation \ref{eq:Rankine-Hugoniot}), give a Chapman-Jouguet detonation velocity of $D_{\mathrm{CJ}} = \sqrt{q} = 1$ and a shock speed that is the average of the pre- and post-shock states $D= \frac{u_{\mathrm{s}}}{2}$ (for $u=0$ in the unshocked gas; the subscript ``$\mathrm{s}$'' denotes the shocked state).

\section{\label{numerical_method}Numerical Method}

The flow field was initiated by an under-driven piston where $u=0.5$ behind the shock, $u=0$ elsewhere, and $\lambda = 0$ everywhere.  The simulations were run for 100 pulse times, but reached their regular oscillatory behaviour much sooner.

\begin{figure}%[h!]
%	\centering
%	\begin{subfigure}[t]{0.66\textwidth}
	\begin{subfigure}[t]{0.5\textwidth}
		\centering
		\includegraphics[scale=1]{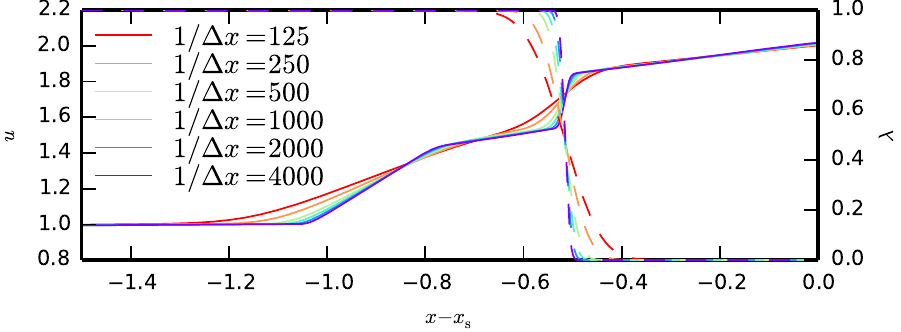}
		%\caption{Profiles of $u$ (solid) and $\lambda$ (dotted)}
		\caption{}
	\end{subfigure}%
\\
%	\begin{subfigure}[t]{0.34\textwidth}
	\begin{subfigure}[t]{0.5\textwidth}
		\centering
		\includegraphics[scale=1]{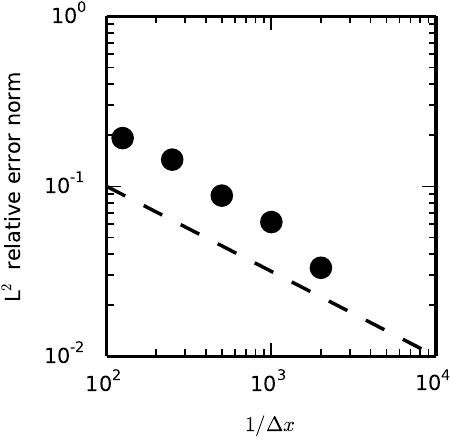}
		%\caption{\label{fig:resolution_study:quantitative}Error v. resolution; dashed line shows $(1/\Delta x)^{-1/2}$}
		\caption{}
		\label{fig:resolution_study:quantitative}
	\end{subfigure}%
	\caption{\label{fig:resolution_study}Effect of resolution on profiles and error at $t=99.5$; (a) profiles of $u$ (solid) and $\lambda$ (dashed), and (b) error vs resolution; the dashed line shows $(1/\Delta x)^{-1/2}$}	
\end{figure}

A uniform grid of 4000 points per $D_{\mathrm{CJ}} t_{\mathrm{p}}$ was used to discretize the domain. The effect of resolution is shown in figure \ref{fig:resolution_study}. Increasing the resolution increased the sharpness of discontinuities, but had no qualitative effect on the phenomena. The L$^2$ relative error norm $\sqrt{\sum(u - u_{\mathrm{ref.}})^2 / \sum (u_{\mathrm{ref.}})^2}$ in figure \ref{fig:resolution_study:quantitative} shows convergence of the solution towards the most-resolved case ($u_{\mathrm{ref.}}$, with $\frac{1}{\Delta x} = 4000$) at the expected rate. %(dashed line).

The Riemann problem was solved at every cell interface using a first-order Godunov method \cite{clarke_numerical_1989} with a first-order Euler method in time. The time step size was the minimum between the time step dictated by the Courants-Freidrich-Lewy (CFL) condition and the time to the next scheduled pulse,
\begin{equation}
\Delta t = \min \left(\mathrm{CFL} \times \frac{\Delta x}{u_{\mathrm{max}}}, t_{\mathrm{p}}- t\bmod{t_{\mathrm{p}}} \right)
\end{equation}
with $\mathrm{CFL} = 0.5$. A source $\ode{\lambda}{t} = \frac{(1-\lambda)}{\Delta t}$ was activated everywhere in the domain when a pulse occurred.

{\color{black}Simulations were performed in the shock-attached frame of reference}. The post-shock state was used as the right boundary condition, and a zero-gradient condition was used on the left side of a domain with a length of $2 D_{\mathrm{CJ}} t_{\mathrm{p}}$. The domain size did not impact the travelling wave solution because a sonic point was formed a distance $D_{\mathrm{CJ}} t_{\mathrm{p}}$ behind the shock, isolating the detonation from the left boundary condition.

Simulations were also performed in the laboratory frame of reference, where the reaction front is stationary and the shock moves relative to the mesh. The laboratory-frame results were qualitatively and quantitatively similar to the shock-fit simulations that will be presented. 

{\color{black}Numerical shock splitting problems \cite{colellaTheoreticalNumericalStructure1986} did not appear because the reaction and shock fronts only coincide for a short period of time, and the shock-fit boundary does not permit numerical diffusion ahead of the shock wave.}

\section{\label{sec:results}Results}

\begin{figure}%[h!]
	\centering
	\includegraphics[scale=1]{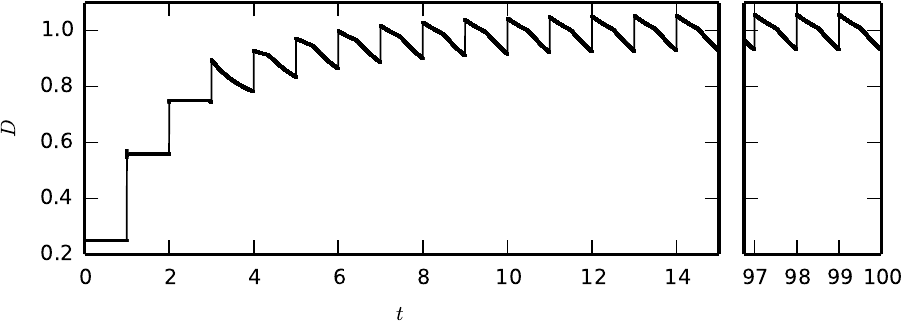}%\vspace{-0.5cm}
	\caption{\label{fig:pulse_detonation_speed}Detonation speed evolution}	
\end{figure}

The time evolution of the shock front speed is plotted in figure \ref{fig:pulse_detonation_speed}. The initial shock speed is maintained until the first pulse at $t=1$. The instantaneous reaction of all shocked material when the pulse occurs causes the sudden acceleration of the shock front. This is repeated at every pulse. A cyclical saw tooth profile is developed by $t=4$.

\begin{figure}%[h!]
	\centering
	%\begin{subfigure}[b]{0.33\textwidth}
	\begin{subfigure}[b]{0.25\textwidth}
		\centering
		\includegraphics[scale=1]{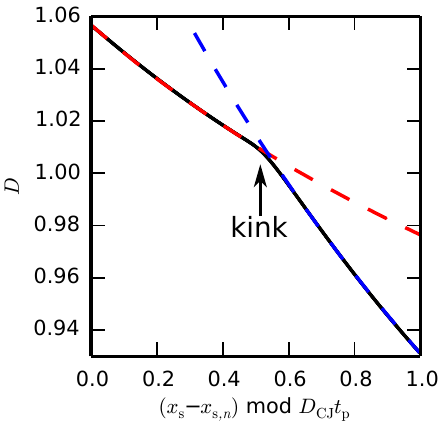}%\vspace{-0.25cm}
		%\caption{$D$ over the cycle}
		\caption{}
		\label{fig:phase_diagrams:D_xs}
	\end{subfigure}%
	%\begin{subfigure}[b]{0.33\textwidth}
	\begin{subfigure}[b]{0.25\textwidth}
		\centering
		\includegraphics[scale=1]{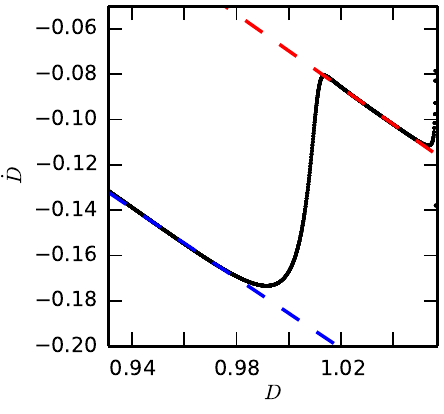}%\vspace{-0.25cm}
		%\caption{$\dot{D}$ and $D$}
		\caption{}
		\label{fig:phase_diagrams:dotD_D}
	\end{subfigure}%
	\\%\hfill%
	\begin{subfigure}[b]{0.33\textwidth}
		\centering
		\includegraphics[scale=1]{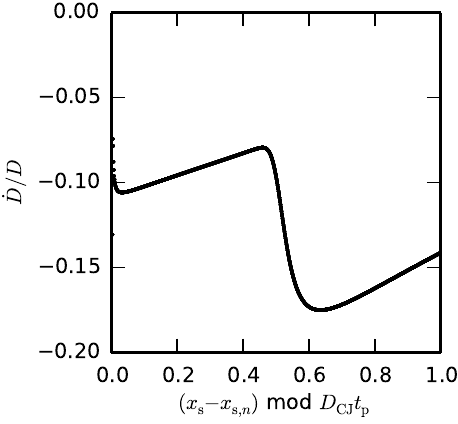}%\vspace{-0.25cm}
		%\caption{Time scale}
		\caption{}
		\label{fig:phase_diagrams:timescale}
	\end{subfigure}%
	\caption{Phase diagrams of ten superimposed cycles (black, $90 \le t \le 100$); (a) $D$ over the cycle, (b) $\dot{D}$ and $D$, and (c) time scale; dashed lines show fits; $x_{\mathrm{s},n}$ is the shock position at the $n = 90$ pulse}
	\label{fig:phase_diagrams}
\end{figure}

The phase diagrams of figure \ref{fig:phase_diagrams} show ten superimposed cycles once the oscillatory behaviour is reached. The front speed is plotted against its position in the cycle in figure \ref{fig:phase_diagrams:D_xs}. It has a cycle-averaged detonation velocity $D_{\mathrm{avg}} = 1 = D_{\mathrm{CJ}}$ equal to the CJ detonation velocity.
Mi and Higgins \cite{mi_influence_2015} found the same averaged velocity in their model based on the Burgers equations, however, Radulescu and Shepherd \cite{radulescu2015dynamics} and Mi \textit{et al.} \cite{miEffectSpatialDiscretization2017} observed average detonation speeds above of the CJ speed in their models using the Euler equations (footnote \footnote{Mi \textit{et al.} \cite{miEffectSpatialDiscretization2017} found that the averaged detonation velocity increasingly exceeded the CJ velocity as the ratio of specific heats was decreased.}). Mi \textit{et al.} \cite{miEffectSpatialDiscretization2017} hypothesized the discrepancy was due to the lack of backwards-traveling characteristics in the Burgers system, but to date, this finding remains unexplained.

The saw tooth velocity profile shown in figure \ref {fig:phase_diagrams:D_xs} has a kink that separates a segment of slow decay from a segment of fast decay. The unsteadiness of detonation waves has previously been modeled \cite{lundstrom1969influence} after Taylor-Sedov blast decay. In similar fashion, the shock speed is fit to a power law
\begin{equation}
\label{eq:power_law_fit}
%D = D_c \left( \frac{x_{\mathrm{s}} - x_0}{x_c - x_0}\right)^{\theta}
D \propto \left( x_{\mathrm{s}} - x_{\mathrm{b}}\right)^{\theta},
%\ln(D) = \theta \ln(x_{\mathrm{s}} - x_0) + C
\end{equation}
where the fitting parameters are the decay exponent $\theta$ and the blast origin $x_{\mathrm{b}}$.
%$D_c$ and $x_{\mathrm{s},c}$ are the shock speed and position at a known point. 
The decay exponents, fit before and after the kink (dashed lines in figure \ref{fig:phase_diagrams:D_xs}, $\theta = -0.143$ in red, $\theta = -0.378$ in blue; footnote \footnote{Both segments decay slower than an inert blast ($\theta = - 1$) would decay. The other fitting parameters are  $x_{\mathrm{b}} = -1.36$ (red) and $x_{\mathrm{b}} = -1.46$ (blue).}), are found to be different from each other despite the system only oscillating at period-one.

Plotted in $\dot{D}$-$D$ phase space, the two segments have similar slopes (dashed lines in figure \ref{fig:phase_diagrams:dotD_D}, $-0.77$ in red, $-0.81$ in blue).
Nevertheless, two distinct groups of time scales are formed, expressed as ${\dot{D}}/{D}$ in figure \ref{fig:phase_diagrams:timescale}, ranging continuously from ${\dot{D}}/{D} = -0.11$ to $-0.074$ at the start of the cycle, and from ${\dot{D}}/{D} = -0.18$ to $-0.14$ at its end. 
This simple model, forced at a single period, recovers two decay exponents and two clearly distinct groups of time scales.

{\color{black}There have been numerous investigations of the one-dimensional dynamics of detonations using more complex systems such as the Euler equations with single-step reactions \cite{ng_nonlinear_2005,henrick_simulations_2006} and losses \cite{sowMeanStructureOnedimensional2014}, the Navier-Stokes equations with one-step \cite{romick2012effect} and full chemistry \cite{romickVerifiedValidatedCalculation2015,hanPulsationOnedimensionalH22019}. In all cases, two distinct groups of decay rates can be found in period-one oscillations \cite{romickVerifiedValidatedCalculation2015}, higher periods \cite{henrick_simulations_2006,sowMeanStructureOnedimensional2014,romickVerifiedValidatedCalculation2015}, and in the chaotic behaviour \cite{romickVerifiedValidatedCalculation2015}. The phase diagram of figure \ref{fig:phase_diagrams:dotD_D} shows a slow decay when the shock is strong and rapid decay when the shock is weak. In the complex models, the fastest decay is observed when the shock is strong. This phase shift might be explained by the lack of re-amplification phase in the simplified model, but further study of the complex systems is required.}

\begin{figure}
	%\begin{minipage}{.5\textwidth}
		\centering
		\begin{subfigure}[b]{0.5\textwidth}
			\centering
			$\mathrel{\raisebox{7.5mm}{$t = 0.00$}}$\hspace{3mm}
			\includegraphics[scale=1]{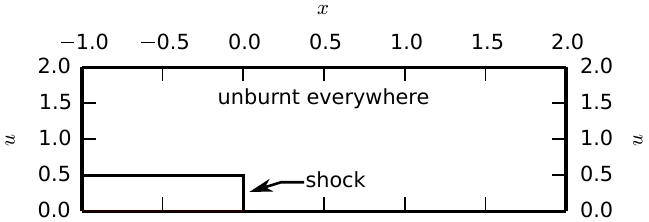}
			\vspace*{-2.mm}%\subcaption{\label{fig:initiation_profile:a}$t = 0$}\vspace*{-0mm}
			\subcaption{\label{fig:initiation_profile:a}}\vspace*{-0mm}
		\end{subfigure}\vspace*{-1.mm}
		\begin{subfigure}[b]{0.5\textwidth}
			\centering
			$\mathrel{\raisebox{7.5mm}{$t = 0.99$}}$\hspace{3mm}
			\includegraphics[scale=1]{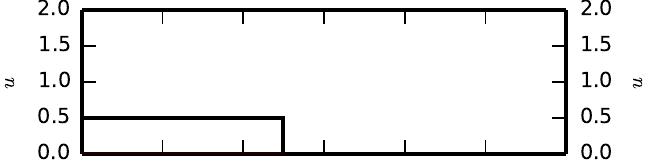}
			\vspace*{-2.mm}%\subcaption{\label{fig:initiation_profile:b}$t = 0.99$}
			\subcaption{\label{fig:initiation_profile:b}}
		\end{subfigure}\vspace*{-1.mm}
		\begin{subfigure}[b]{0.5\textwidth}
			\centering
			$\mathrel{\raisebox{7.5mm}{$t = 1.01$}}$\hspace{3mm}
			\includegraphics[scale=1]{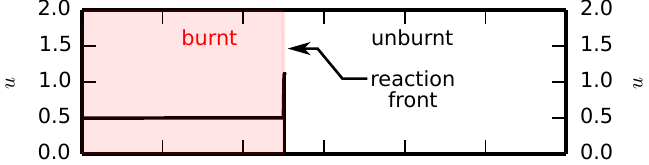}
			\vspace*{-2.mm}%\subcaption{\label{fig:initiation_profile:c}$t = 1.01$}
			\subcaption{\label{fig:initiation_profile:c}}
		\end{subfigure}\vspace*{-1.mm}
		\begin{subfigure}[b]{0.5\textwidth}
			\centering
			$\mathrel{\raisebox{7.5mm}{$t = 1.99$}}$\hspace{3mm}
			\includegraphics[scale=1]{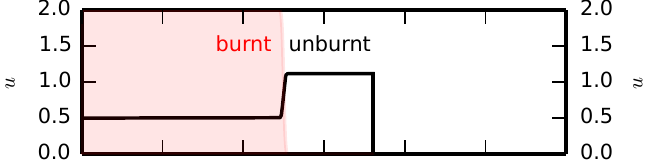}
			\vspace*{-2.mm}%\subcaption{\label{fig:initiation_profile:d}$t = 1.99$}
			\subcaption{\label{fig:initiation_profile:d}}
		\end{subfigure}\vspace*{-1.mm}
		\begin{subfigure}[b]{0.5\textwidth}
			\centering
			$\mathrel{\raisebox{7.5mm}{$t = 2.01$}}$\hspace{3mm}
			\includegraphics[scale=1]{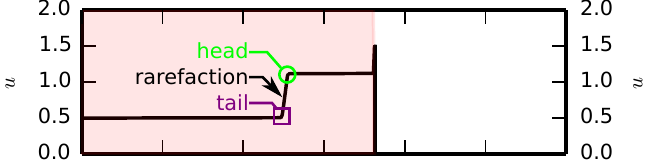}
			\vspace*{-2.mm}%\subcaption{\label{fig:initiation_profile:e}$t = 2.01$}
			\subcaption{\label{fig:initiation_profile:e}}
		\end{subfigure}\vspace*{-1.mm}
		\begin{subfigure}[b]{0.5\textwidth}
			\centering
			$\mathrel{\raisebox{7.5mm}{$t = 2.20$}}$\hspace{3mm}
			\includegraphics[scale=1]{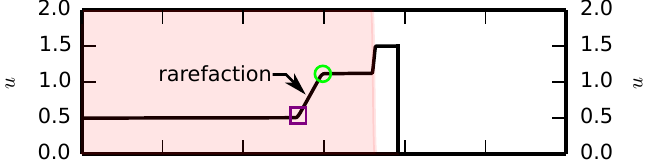}
			\vspace*{-2.mm}%\subcaption{\label{fig:initiation_profile:f}$t = 2.2$}
			\subcaption{\label{fig:initiation_profile:f}}
		\end{subfigure}\vspace*{-1.mm}
		\begin{subfigure}[b]{0.5\textwidth}
			\centering		
			$\mathrel{\raisebox{7.5mm}{$t = 2.40$}}$\hspace{3mm}
			\includegraphics[scale=1]{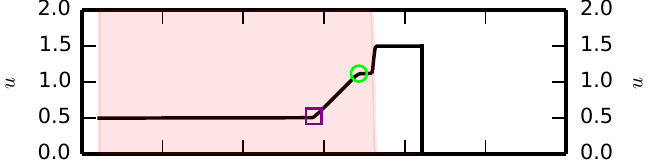}
			\vspace*{-2.mm}%\subcaption{\label{fig:initiation_profile:g}$t = 2.4$}
			\subcaption{\label{fig:initiation_profile:g}}
		\end{subfigure}\vspace*{-1.mm}
		\begin{subfigure}[b]{0.5\textwidth}	
			\centering
			$\mathrel{\raisebox{7.5mm}{$t = 2.60$}}$\hspace{3mm}
			\includegraphics[scale=1]{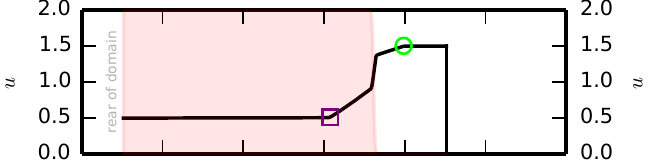}
			\vspace*{-2.mm}%\subcaption{\label{fig:initiation_profile:h}$t = 2.6$}
			\subcaption{\label{fig:initiation_profile:h}}
		\end{subfigure}\vspace*{-1.mm}
		\begin{subfigure}[b]{0.5\textwidth}
			\centering
			$\mathrel{\raisebox{7.5mm}{$t = 2.99$}}$\hspace{3mm}
			\includegraphics[scale=1]{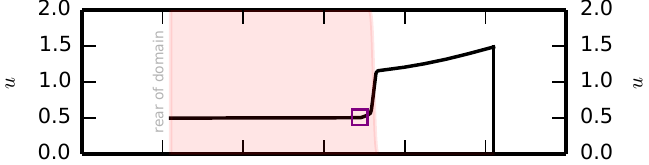}
			\vspace*{-2.mm}%\subcaption{\label{fig:initiation_profile:i}$t = 2.99$}
			\subcaption{\label{fig:initiation_profile:i}}
		\end{subfigure}\vspace*{-1.mm}
		\begin{subfigure}[b]{0.5\textwidth}
			\centering
			$\mathrel{\raisebox{14.mm}{$t = 3.01$}}$\hspace{3mm}
			\includegraphics[scale=1]{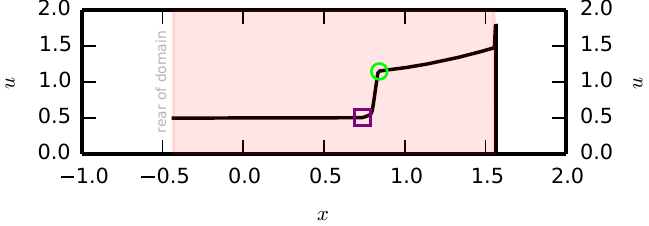}
			\vspace*{-2.mm}%\subcaption{\label{fig:initiation_profile:j}$t = 3.01$}
			\subcaption{\label{fig:initiation_profile:j}}
		\end{subfigure}
	%\end{minipage}
	\caption{Profiles of $u$ (solid, left axis) and the reaction front (at the shaded/unshaded interface) in the laboratory frame of reference during the initial cycles; expansion fan head (circle) and expansion fan tail (square)}
	\label{fig:initiation_profile}
\end{figure}

\begin{figure*}
	\centering
	\begin{subfigure}[t]{0.2\textwidth}
		\centering
		\hspace{5mm}${t_n = n}$ %\vspace{-1mm}
		\includegraphics[scale=1]{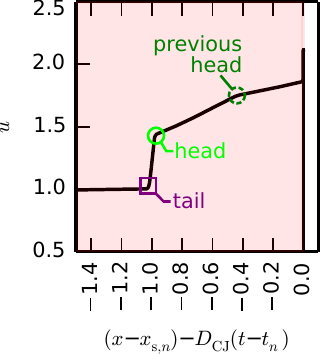}
		%\caption{\label{fig:pulse_profiles_steady:a}${t_n = n}$}
		\caption{\label{fig:pulse_profiles_steady:a}}
	\end{subfigure}%
	\begin{subfigure}[t]{0.15\textwidth}
		\centering
		${t_n + 0.2}$
		\includegraphics[scale=1]{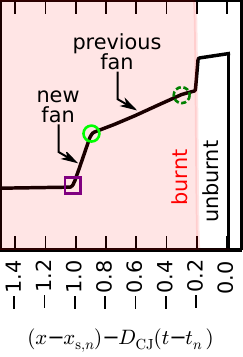}
		%\caption{\label{fig:pulse_profiles_steady:b}${t_n + 0.2}$}
		\caption{\label{fig:pulse_profiles_steady:b}}
	\end{subfigure}%
	\begin{subfigure}[t]{0.15\textwidth}
		\centering
		${t_n + 0.4}$
		\includegraphics[scale=1]{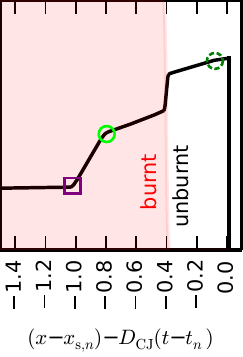}
		%\caption{\label{fig:pulse_profiles_steady:c}${t_n + 0.4}$}
		\caption{\label{fig:pulse_profiles_steady:c}}
	\end{subfigure}%
	\begin{subfigure}[t]{0.15\textwidth}
		\centering
		${t_n + 0.6}$
		\includegraphics[scale=1]{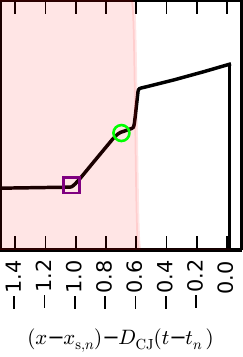}
		%\caption{\label{fig:pulse_profiles_steady:d}${t_n + 0.6}$}
		\caption{\label{fig:pulse_profiles_steady:d}}
	\end{subfigure}%
	\begin{subfigure}[t]{0.15\textwidth}
		\centering
		${t_n + 0.8}$
		\includegraphics[scale=1]{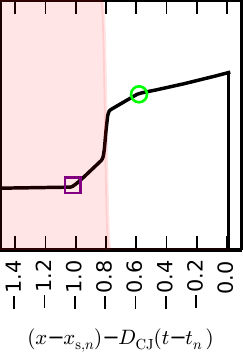}
		%\caption{\label{fig:pulse_profiles_steady:e}${t_n + 0.8}$}
		\caption{\label{fig:pulse_profiles_steady:e}}
	\end{subfigure}%
	\begin{subfigure}[t]{0.2\textwidth}
		\centering
		\hspace{-5mm}${t_n + 0.99}$
		\includegraphics[scale=1]{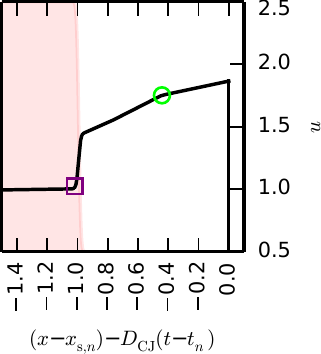}
		%\caption{\label{fig:pulse_profiles_steady:f}${t_n + 0.99}$}
		\caption{\label{fig:pulse_profiles_steady:f}}
	\end{subfigure}%
	\caption{Profiles of $u$ (solid, left axis) and the reaction front  (at the shaded/unshaded interface) over one regular cycle in a frame of reference moving at $D_{\mathrm{CJ}}$; expansion fan heads from current (solid circle) and previous (dashed circle) cycles, and expansion fan tail (square); ($n=99$)}
	\label{fig:pulse_profiles_steady}
\end{figure*}

The initial transient is plotted in figure \ref{fig:initiation_profile} in the laboratory frame of reference through snapshots of $u$ and the reaction front position. %, and for the fully developed pulsations in figure \ref{fig:pulse_profiles_steady} will be examined next to determine the origin of the two distinct time scales.
The uniform initial conditions (figure \ref{fig:initiation_profile:a}) of $u=0.5$ behind the shock initially maintain the constant shock speed (figure \ref{fig:initiation_profile:b}). When the first pulse occurs at $t=1$ (figure \ref{fig:initiation_profile:c}), the reaction front is moved to the shock front, strengthening it. The shock travels faster than before the pulse, leaving the motionless reaction front behind (figure \ref{fig:initiation_profile:d}).
There is a jump in $u$ across the reaction front, with a higher value of $u$ in the reactants than in the products.  
%The second pulse occurs (figure \ref{fig:initiation_profile:e}), bringing the reaction front to the shock. The shock speed is immediately increased, and the shock once again leaves the new reaction front behind. 
The same events are repeated for the second pulse (figure \ref{fig:initiation_profile:e}). 
Additionally, an expansion wave is created at the reaction front's pre-pulse location because the discontinuity in $u$ is no longer supported by the reaction front (figure \ref{fig:initiation_profile:f}). The head of this expansion fan (the fastest portion of the rarefaction, circled in green) travels towards the shock front and is amplified as it crosses through the reaction discontinuity (figure \ref{fig:initiation_profile:h}). The head of the rarefaction reaches the shock just before the third pulse and begins to attenuate the shock (figure \ref{fig:initiation_profile:i}).
The same events are repeated at the third pulse (figure \ref{fig:initiation_profile:j}), however, the shock now immediately decays because it is attenuated by the previous pulse's rarefaction. A regular oscillating cycle is soon formed.

The regular cycle is shown in figure \ref{fig:pulse_profiles_steady}, plotted in the frame of reference moving at $D_{\mathrm{CJ}}$ with the origin located at the shock at the beginning of the $n^{\mathrm{th}}$ cycle. The shock front is located on the right side and the reaction front moves to the left at a speed $D_{\mathrm{CJ}}$ in this frame of reference.

The $n^{\mathrm{th}}$ cycle starts at a pulse, with the reaction front at the shock (figure \ref{fig:pulse_profiles_steady:a}). % where the pulse amplifies the shock. 
A rarefaction wave forms at  $(x-x_{\mathrm{s},n}) -D_{\mathrm{CJ}} (t - t_n) = -1$, where the reaction front was located prior to the pulse.
The head of the rarefaction (solid circle) travels to the right faster than its tail (square) and the shock speed. The speed disparity between the head and tail forms the expansion fan, \textit{i.e.} the sloped segment seen between the square and circle in figure \ref{fig:pulse_profiles_steady:b}. The fan originates when the jump in $u$ that was supported by the reaction front is abandoned at the pulse because the reaction front moves to the shock after a pulse. % and falls to the left of the shock. % while the freshly amplified shock begins to decay. 

As time passes (figures \ref{fig:pulse_profiles_steady:b}, \ref{fig:pulse_profiles_steady:c} and \ref{fig:pulse_profiles_steady:d}), the spread between the head and tail of the rarefaction grows, the reaction front falls further to the left of the shock, and the shock strength $u_{\mathrm{s}}$ diminishes.
The rarefaction head is amplified as it crosses the reaction front (figure \ref{fig:pulse_profiles_steady:e}).
%The head from the previous cycle reaches the front after $t=14.4$ and the new expansion head (solid circle) moves forward. The head from the current cycle passes through the reaction zone after $t=14.6$.
The cycle terminates (figure \ref{fig:pulse_profiles_steady:f}) with the reaction front at the rear, supporting the discontinuity in $u$ that will become the next cycle's expansion wave. In the next cycle (returning to figure \ref{fig:pulse_profiles_steady:a}), the head of the previous expansion (dotted circle) continues to propagate to the right and is amplified once more (figure \ref{fig:pulse_profiles_steady:c}) before reaching the shock. The head brings a change in the slope of $u$, causing the shock to decay at a new rate.

%\begin{figure}%[h!]
%	\centering
%	\begin{subfigure}[t]{0.17\textwidth}
%		\centering
%		\includegraphics[scale=1]{{./figures/profile/u,L_t=99.01}.pdf}
%		\caption{\label{fig:pulse_profiles_steady:a}${t_0 = n}$}
%	\end{subfigure}\\
%	\begin{subfigure}[t]{0.16\textwidth}
%		\centering
%		\includegraphics[scale=1]{{./figures/profile/u,L_t=99.2}.pdf}
%		\caption{\label{fig:pulse_profiles_steady:b}${t_0 + 0.2}$}
%	\end{subfigure}\\%
%	\begin{subfigure}[t]{0.16\textwidth}
%		\centering
%		\includegraphics[scale=1]{{./figures/profile/u,L_t=99.4}.pdf}
%		\caption{\label{fig:pulse_profiles_steady:c}${t_0 + 0.4}$}
%	\end{subfigure}\\%
%	\begin{subfigure}[t]{0.16\textwidth}
%		\centering
%		\includegraphics[scale=1]{{./figures/profile/u,L_t=99.6}.pdf}
%		\caption{\label{fig:pulse_profiles_steady:d}${t_0 + 0.6}$}
%	\end{subfigure}\\%
%	\begin{subfigure}[t]{0.16\textwidth}
%		\centering
%		\includegraphics[scale=1]{{./figures/profile/u,L_t=99.8}.pdf}
%		\caption{\label{fig:pulse_profiles_steady:e}${t_0 + 0.8}$}
%	\end{subfigure}\\ %
%	\begin{subfigure}[t]{0.17\textwidth}
%		\centering
%		\includegraphics[scale=1]{{./figures/profile/u,L_t=99.99}.pdf}
%		\caption{\label{fig:pulse_profiles_steady:f}${t_0 + 0.99}$}
%	\end{subfigure}%
%	\caption{Profiles of $u$ (solid, left axis) and the reaction front (dotted, right axis) over one regular cycle in a frame of reference moving at $D_{\mathrm{CJ}}$; expansion fan heads from current (solid circle) and previous (dashed circle) cycles, and expansion fan tail (square)}
%	\label{fig:pulse_profiles_steady}
%\end{figure}

The periodic creation of a rarefaction when the reaction front moves to 
%\begin{wrapfigure}{r}{0.47\textwidth}
\begin{figure}
	%\vspace{0.cm}\hspace{-0.5cm}
	\includegraphics[scale=1]{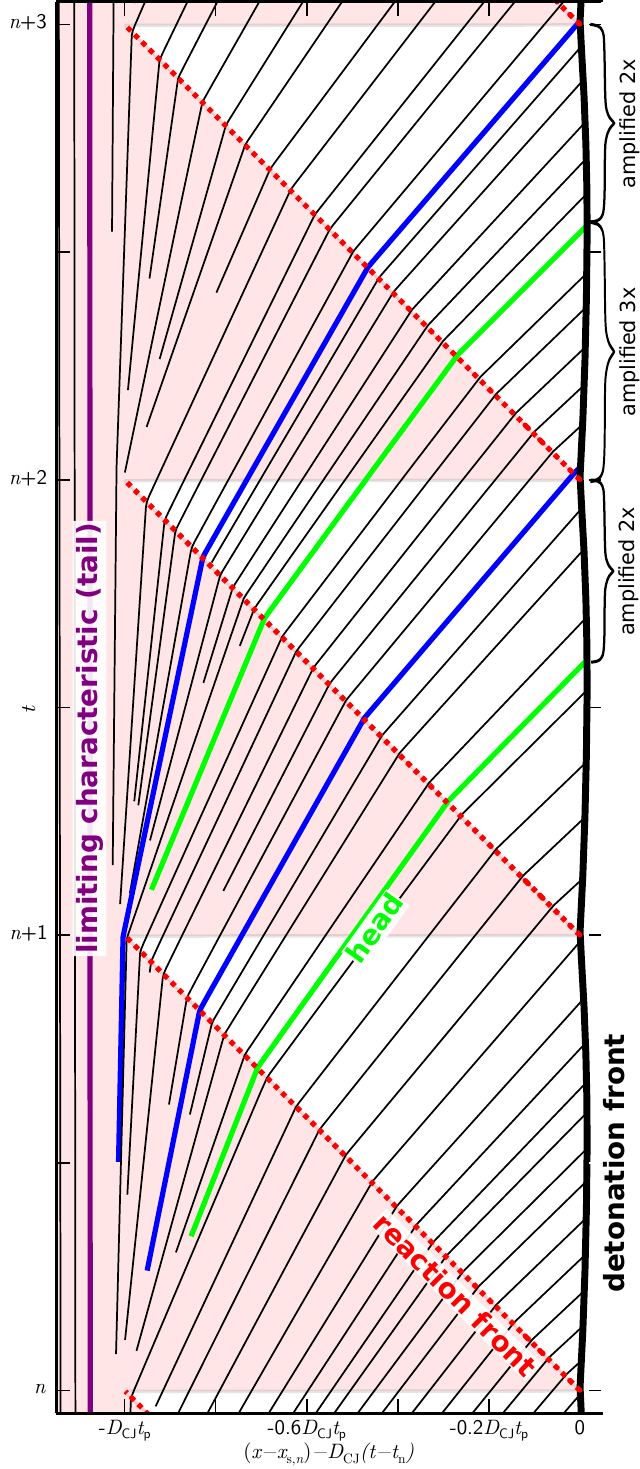}%
	\caption{\label{fig:characteristics}Characteristics diagram (black) and reaction front position (red dotted line, reacted where shaded) over three periods of the regular pulsating behaviour; blue and green characteristics delimit characteristics amplified twice from those amplified three times ($n=97$)}%\vspace{-1.5cm}
\end{figure}
%\end{wrapfigure}
the shock drives the dynamics of the system. The  shock strength is amplified every pulse, then attenuated by the arrival of the expansion wave at the front. %This is repeated every pulse, leading to a regular oscillating cycle.

%In summary, a rarefaction is created every cycle when the reaction front abandons the jump in $u$ to the rear and moves to the shock front. It takes more than one cycle for the head of this rarefaction to reach the shock front. During this time, it is amplified twice across the reaction front. Once it reaches the front, the change in slope across the rarefaction head changes the detonation's decay rate.

\section{\label{sec:analysis}Analysis}

The flow can also be visualized by looking at a characteristic diagram, shown in figure \ref{fig:characteristics} once the regular oscillatory behaviour is reached. The characteristics are plotted (thin black lines) in the frame of reference moving at $D_{\mathrm{CJ}}$. %a steady frame of reference traveling at $D_{\mathrm{CJ}}$. % with the origin situated at the shock front immediately after a pulse. 
The characteristics have slopes of $\mathrm{d}t/ \mathrm{d} x = 1/(u-D_{\mathrm{CJ}})$. %$\ode{t}{x} = 1/(u - D_{\mathrm{CJ}})$.
Characteristics to the left of the limiting characteristic travel slower than the average detonation speed and will never reach the detonation front, whereas characteristics to the right of the limiting characteristic eventually reach the front. The reaction fronts (dotted red line segments) are angled to the left with slope $-1/D_{\mathrm{CJ}}$ in this frame of reference. Characteristics ahead of the detonation are omitted for clarity, but would form lines parallel to the reaction fronts. 

The expansion fan created at the beginning of each pulse lies between the limiting characteristic and the characteristic highlighted in green, centered at ${t = t_n=n }$ and ${(x-x_{\mathrm{s},n}) -D_{\mathrm{CJ}} (t - t_n) = -D_{\mathrm{CJ}} t_{\mathrm{p}}}$, where $x_{\mathrm{s},n}$ is the shock location at the $n^{\mathrm{th}}$ pulse. The green characteristic represents the head of the rarefaction, the fastest characteristic in the expansion fan. The characteristics in the fan have constant slope (\textit{i.e.} constant $u$) until they cross the reaction front, where they are amplified and accelerate, seen by their change in slope. 
The characteristics are amplified a second time as they pass through the reaction front from the next pulse. The head characteristic reaches the shock front around $t \approx n+1.6$. However, not all of the characteristics from this fan reach the front before the next pulse. The last characteristic to reach the front before the $n+2$ pulse is highlighted in blue. All characteristics above the blue characteristic are amplified once more before reaching the shock, more than two periods after they were released.

Characteristics that are only amplified twice cause faster decay than those amplified three times, leading to the two distinct decay exponents of the shock speed. This time scale separation of forward-facing pressure waves may be responsible for the period-doubling behaviour of detonations.

The strength of the characteristics, times, and locations of their amplifications can be found analytically thanks to the simplicity of the model. An expansion fan is created at every pulse at distance $D_{\mathrm{CJ}} t_{\mathrm{p}}$ behind the detonation front. From here, each characteristic travels at a constant speed towards the detonation front until it is amplified across the reaction front. The amount of amplification across the reaction front is given by the Rankine-Hugoniot jump condition
\begin{equation}
\label{eq:Rankine-Hugoniot}
S = \frac{\left[\frac{1}{2}u^2 + \frac{1}{2} q \lambda\right]}{\left[u\right]} = \frac{\left(\frac{1}{2}u_{\mathrm{r}}^2 + \frac{1}{2} q \lambda_{\mathrm{r}} \right) - \left(\frac{1}{2}u_{\mathrm{l}}^2 + \frac{1}{2} q \lambda_{\mathrm{l}} \right)}{\left(u_{\mathrm{r}} - u_{\mathrm{l}}\right)}
\end{equation}
for a discontinuity with speed $S$ (footnote \footnote{Solving the Rankine-Hugoniot equation (\ref{eq:Rankine-Hugoniot}) for a CJ detonation yields $S = D_{\mathrm{CJ}} =\sqrt{q}$ using the sonic condition $u_{\mathrm{l}} = D_{\mathrm{CJ}}$ (and $\lambda_{\mathrm{l}}=1$, $\lambda_{\mathrm{r}}=0$, $u_{\mathrm{r}}=0$); for a shock propagating into $u_{\mathrm{r}}=0$ it yields $S = \frac{u_{\mathrm{s}}}{2}$, where $u_{\mathrm{s}}$ is the post-shock state (and $\lambda_{\mathrm{l}}=\lambda_{\mathrm{r}}=0$, $u_{\mathrm{l}} = u_{\mathrm{s}}$, $u_{\mathrm{r}}=0$).}). Subscripts l and r denote the states to the left and right of the discontinuity. Since the reaction front is immobile in the laboratory frame of reference ($S=0$), fully reacted to its left ($\lambda_{\mathrm{l}}=1$) and unreacted to its right ($\lambda_{\mathrm{r}}=0$), the equation simplifies to
\begin{equation}
\label{eq:amplification}
u_{\mathrm{r}} = \sqrt{u_{\mathrm{l}}^2 + q}.
\end{equation}
Given the strength $u_{\mathrm{l}}$ of a characteristic that enters the reaction front from the left side, its amplified strength $u_{\mathrm{r}}$ on the right is known.
This means the expansion fan created at each pulse ranges from $u=D_{\mathrm{CJ}}$ at the tail of the fan, where the limiting characteristic is unamplified, to $u = \sqrt{D_{\mathrm{CJ}}^2 + q}$ at the head where it is the strongest. The strength of intermediate characteristics in the expansion fan, before their amplification at $x=x_{\mathrm{s},n}$, is given by
\begin{equation}
u=\frac{x - x_{\mathrm{f}}}{t - t_{\mathrm{f}}}
\end{equation}
where $(x_{\mathrm{f}}, t_{\mathrm{f}})$ is the center of the fan.

At the $n^{\mathrm{th}}$ pulse, a characteristic of strength $u_0$ travels towards the shock from its birth place $({x_0 = x_{\mathrm{s},n}-D_{\mathrm{CJ}} t_{\mathrm{p}}}, \hspace{0.3cm} t_0 = n)$.
Along this characteristic,
%\begin{equation}
$	t = \frac{1}{u_0} (x - x_0) + t_0$ %\mathrm{\ for\ } x_0 \le x \le x_1.
%\end{equation}
until it reaches the reaction front at $x = x_1 = x_{\mathrm{s},n}$ and the characteristic is amplified to $u_1 = \sqrt{u_0^2 + q}$. The procedure is repeated until the characteristic reaches the shock front. Generally, the $k^{\mathrm{th}}$ intersection between a characteristic and a reaction front occurs at the point $(x_k, t_k)$
\begin{equation}
\label{eq:intersection}
x_k = x_{\mathrm{s},n} + (k-1) D_{\mathrm{CJ}} t_{\mathrm{p}}, 
\mathrm{\hspace{0.5cm}} 
t_k = t_0 + \sum_{i=0}^{k-1} \frac{D_{\mathrm{CJ}} t_{\mathrm{p}}}{\sqrt{u_0^2 + i q}}
\end{equation}
and amplifies the characteristic from $u_0$ to
\begin{equation}
\label{eq:amplification:multiple}
u_k =\sqrt{u_0^2 + kq}.
\end{equation}
Between the $k^{\mathrm{th}}$ and $k+1$ intersections, the characteristic follows
\begin{equation}
\label{eq:between_intersections}
t = \frac{1}{u_k} (x - x_k) + t_k \mathrm{\hspace{1cm} for \hspace{1cm}} x_k \le x \le x_{k+1}.
\end{equation}

This procedure can be used to find the minimum and maximum shock speed, for example. Consider the blue characteristic which arrives at the shock exactly at a pulse. It is the characteristic's third intersection with the pulse ($k=3$) and the intersection occurs two pulses after its formation ($t_k - t_0 = 2 t_{\mathrm{p}}$).
Substituting $u_0$ from equation \ref{eq:amplification:multiple} into the time equation of (\ref{eq:intersection}) gives
\begin{equation}
2 t_{\mathrm{p}} = \frac{D_{\mathrm{CJ}} t_{\mathrm{p}}}{\sqrt{u_3^2 - 3 q}} + \frac{D_{\mathrm{CJ}} t_{\mathrm{p}}}{\sqrt{u_3^2 - 2 q}} + \frac{D_{\mathrm{CJ}} t_{\mathrm{p}}}{\sqrt{u_3^2 - q}} 
\end{equation}
and solving numerically for $u_3$ yields the characteristic strength $u_3 = 2.113$ at the shock immediately after the pulse. Equation \ref{eq:amplification} is used to find its strength immediately before the pulse, $u_2 = 1.862$.
The shock speed calculated from the Rankine-Hugoniot relation (equation \ref{eq:Rankine-Hugoniot}) yields ${D_{\mathrm{min}} = \frac{u_2}{2} = 0.931}$ and ${D_{\mathrm{max}} = \frac{u_3}{2} = 1.057}$. This agrees with the values obtained from simulations. 

Now consider the time at which the green characteristic (the head of the rarefaction) reaches the shock front. 
The head characteristic initially has the strength of the limiting characteristic amplified once, $u_0 = \sqrt{D_{\mathrm{CJ}}^2 + q} = \sqrt{2}$, and travels along the path given by equation \ref{eq:between_intersections} until it reaches the shock, after being amplified twice ($k=2$).
Unfortunately there is no analytical expression for the shock position. It can be integrated numerically but, for simplicity, assume the shock travels at a steady velocity $D_{\mathrm{CJ}}$ because the detonation deviates little from the CJ velocity, as evidenced in figures \ref{fig:phase_diagrams:D_xs} and \ref{fig:characteristics}. With this assumption, the shock path follows $t \approx \frac{1}{D_{\mathrm{CJ}}}(x - x_{\mathrm{s},n}) + t_{\mathrm{0}}$. The intersection of the two paths occurs
\begin{equation}
t \approx t_0 + D_{\mathrm{CJ}} t_{\mathrm{p}} \frac{\left( \frac{1}{\sqrt{u_0^2}} + \frac{1}{\sqrt{u_0^2 + q}} - \frac{1}{\sqrt{u_0^2 + 2 q }} \right)}{1 - \frac{D_{\mathrm{CJ}}}{\sqrt{u_0^2 + 2q}}} = 1.57
\end{equation}
pulses after the head characteristic is born, which agrees with simulations. The analytic description accurately represents the simplified detonation model.

\section{\label{sec:conclusion}Conclusion}

A simplified pulsating detonation model was studied using the Fickett-Majda asymptotic model for detonations in the limit of small heat release and the Newtonian limit; a periodic reaction that instantaneously consumed all shocked gases was implemented. The simple system was used to study the decay behaviour of pulsating detonations. The detonation was found to travel at a pulse-averaged speed equal to the Chapman-Jouguet velocity. The shock speed decay between pulses was fit to a power law. Two decay exponents were found for each oscillation due to a kink in shock speed. This is accompanied by the presence of two distinct groups of time scales every cycle, a feature present in period-two detonations.

After each pulse, a strong expansion wave is created at the last location of reaction front. A characteristic investigation revealed that characteristics originating from the head of this expansion take approximately 1.6 periods to reach and attenuate the detonation front, while characteristics from the tail take an additional period. The leading characteristics are amplified twice by passing through subsequent reaction fronts, before arriving at the shock, while the initially weaker trailing characteristics are amplified three times. These dynamics produce a kinked velocity profile with two groups of time scales. The two sets of time scales intrinsic to pulsating detonations may be the source of period-doubling bifurcations which lead to chaos in more complicated systems. {\color{black}Further study is required to see how the dynamics described in this study may contribute to the two distinct decay rates also seen in more complex systems.}

\section{\label{sec:Acknowledgements}Acknowledgements}

The authors thank Drs. Henrick, Aslam and Powers for the permission to use figure \ref{fig:Henrick_et_al_2006_fig7}, and Drs. Kasimov, Faria and Rosales for the permission to use figure \ref{fig:Kasimov_2013_fig2}. 

\bibliography{apsguide4-2.bib}

\end{document}